\documentclass[twocolumn, prl, superscriptaddress]{revtex4-1}

\usepackage{amsmath}
\usepackage{amssymb}
\usepackage{braket}
\usepackage{bbold}

\usepackage[utf8]{inputenc}

\usepackage[colorlinks,bookmarks=false,linkcolor=black,urlcolor=black,citecolor=black]{hyperref}

\usepackage{graphicx}

\usepackage{epstopdf}

\def \show#1 {{\begin{minipage}{2cm}
      \includegraphics[width=2cm]{schematics#1}
    \end{minipage}}}

\usepackage{array}

\usepackage[caption=false]{subfig}
\newcommand{\SUBFIG}[1]{\subfloat[]{#1}}

\newcommand{\OP}[1]{\hat \psi _{#1}}
\newcommand{\OPd}[1]{\hat \psi _{#1} ^\dagger}

\newcommand{\dDO}[1]{\hat {\delta \rho_{#1}}}
\newcommand{\dPO}[1]{\hat {\delta \varphi_{#1}}}

\newcommand{\OPz}[1]{\Psi_{#1}}

\newcommand{\derx}{\partial_x}

\newcommand{\MAT}[1]{\begin{pmatrix} #1 \end{pmatrix}}
\newcommand{\ve}[1]{ {\mathbf #1}}

\date{\today}

\begin{document} \title{Unstable Avoided Crossing in Coupled Spinor Condensates}

\author{Nathan R. Bernier}
\affiliation{Department of Physics, Boston University, Boston MA 02215}
\author{Emanuele G. Dalla Torre}
\author{Eugene Demler}
\affiliation{Department of Physics, Harvard University, Cambridge MA 20138}
\begin{abstract}
    We consider the dynamics of a Bose-Einstein condensate with two internal states, coupled through a coherent drive. 
We focus on a specific quench protocol, in which the sign of the coupling field is suddenly changed. 
    At a mean-field level, the system is transferred from a minimum to a maximum of the coupling energy and can remain dynamically stable, in spite of the development of negative-frequency modes. 
    In the presence of a non-zero detuning between the two states, the ``charge'' and ``spin'' modes couple, giving rise to an  unstable avoided crossing. 
    This phenomenon is generic to systems with two dispersing modes away from equilibrium and constitutes an example of class-$I_o$ non-equilibrium pattern formation in quantum systems.
\end{abstract}
\maketitle

\paragraph*{} 
Non-equilibrium conditions often lead to the spontaneous formations of  regular patterns. This effect can be predicted by the appearance of dynamical instabilities at finite momentum, finite frequency, or both. 
For classical dissipative systems, several types of pattern formations were reviewed and classified by Cross and Hohenberg \cite{crosshohenberg} (see Table \ref{tab:classification}). With the advent of isolated quantum systems, and in particular ultracold atoms, it is natural to  inquire  whether this phenomenon occurs even in the absence of a dissipative bath. 
Examples of pattern formation in unitary quantum system were indeed observed in the roton softening of dipolar condensates \cite{santos03} and in the dynamics of spinor condensates \cite{sadler06,vanbijnen07,Cherng08}. 
They can be described as mode-softening effects, in which the energy of a particular mode is reduced by changing the parameters of the system, until it reaches zero and the mode becomes unstable. 
This mechanism leads to unstable modes with finite momentum and vanishing real frequency  and to the formation of a static pattern (class $I_s$).

\begin{figure}[t]
  \begin{center}
    \SUBFIG{
    \includegraphics[scale=0.3]{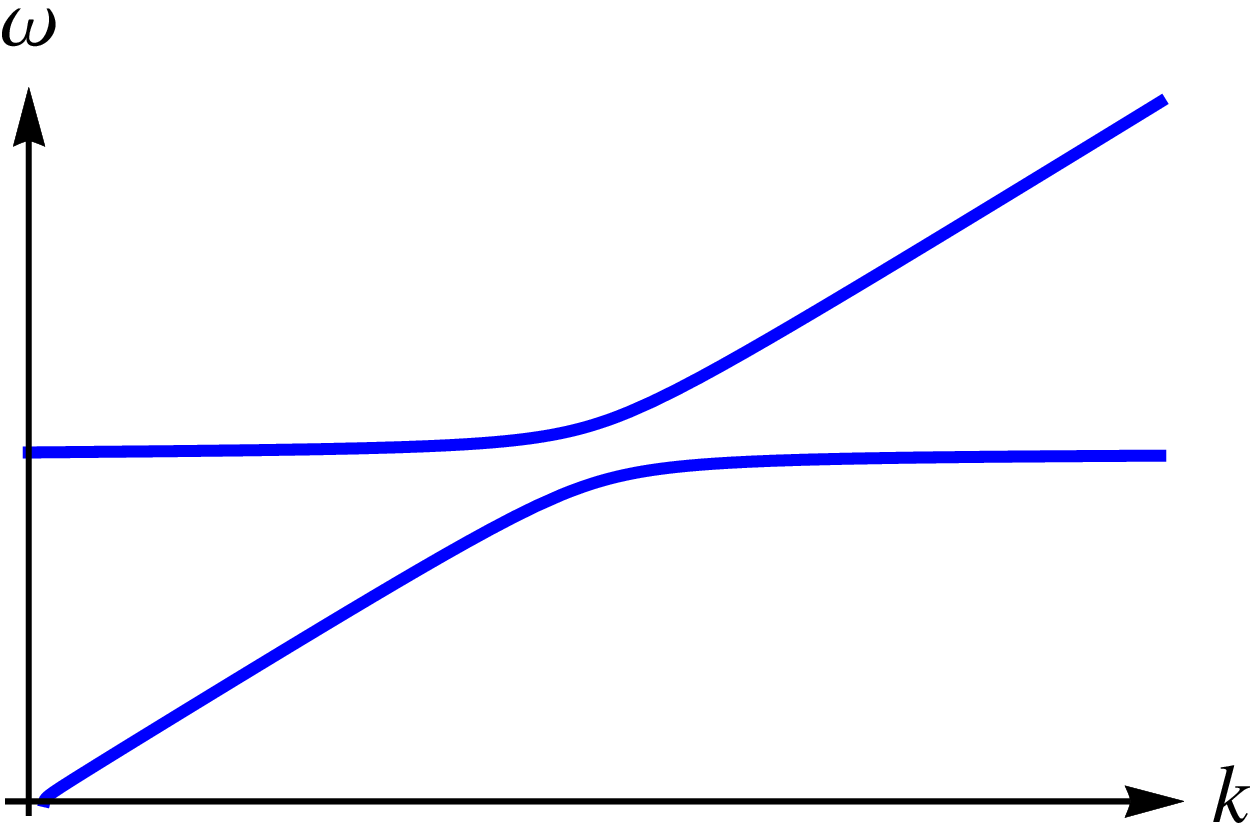}
    \label{subfig:ESES}}
    \SUBFIG{
    \includegraphics[scale=0.3]{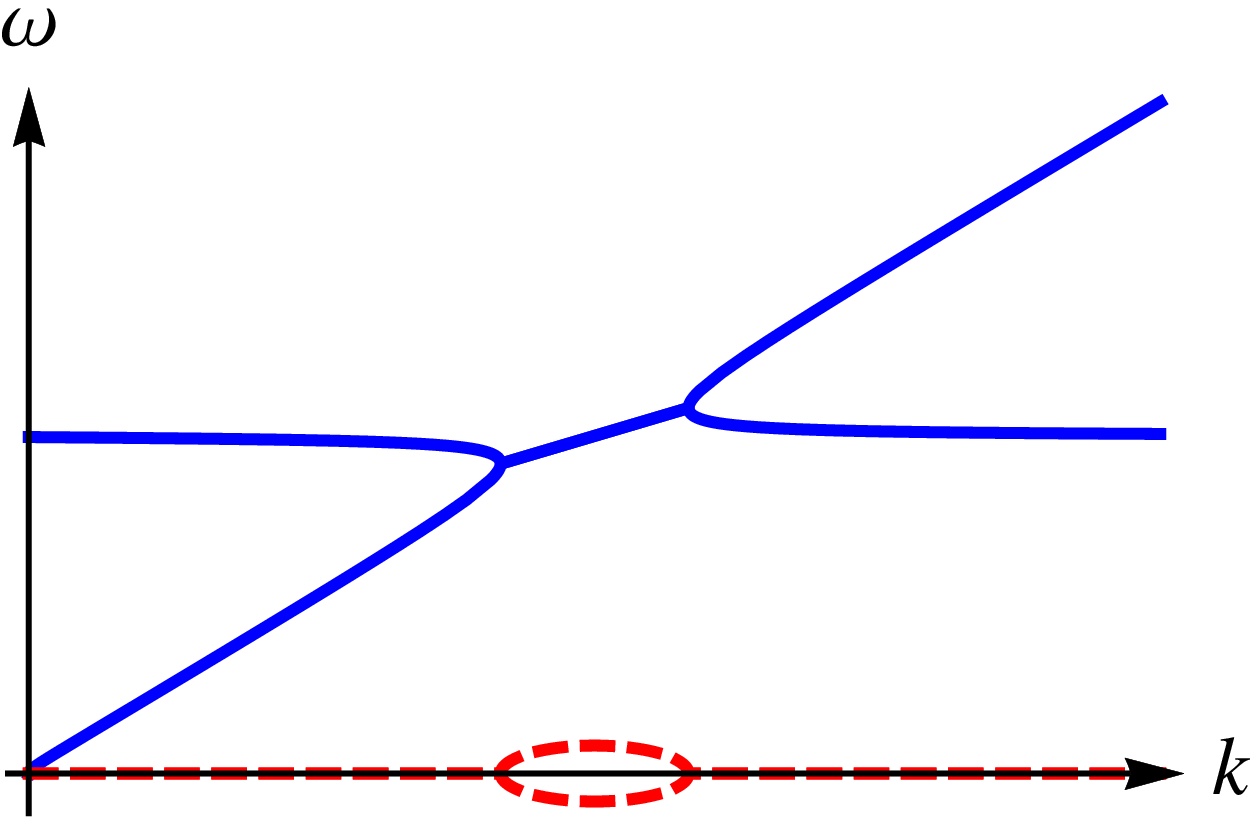}
    \label{subfig:ESDS}}
  \end{center}
  \caption{(color online).
  A toy model demonstration of the stable \protect\subref{subfig:ESES} and unstable \protect\subref{subfig:ESDS} avoided level crossings. 
  One stable mode with a linear dispersion ($\omega_1\propto k$) is mixed with a mode of constant frequency ($\omega_2 = \mathrm{const}$). 
  In \protect\subref{subfig:ESES}, both modes are energetically stable ($\omega_1,\omega_2>0$); in \protect\subref{subfig:ESDS}, the second mode is only dynamically stable ($\omega_2<0$).  
  The real part of the diagonalized modes frequency is the solid blue line (chosen to be positive by convention) and the imaginary part the dashed red line. 
  }
  \label{fig:levelavoidance}\vspace{-0.5cm}
\end{figure}

In this letter, we describe a generic mechanism for the formation of oscillatory patterns (class $I_o$) in unitary quantum systems, with a finite wave vector {\em and} a finite real frequency. As we explain in detail below, this type of instability occurs every time a dynamically stable mode crosses an energetically stable one (see Fig.~\ref{fig:levelavoidance}). For example, this situation can be realized in tunneling-coupled condensates \cite{Hipolito10} and antiferromagnetic spinor condensates \cite{Kronjager10}. To observe it in a simple setting, we consider spinor condensates with an external coherent drive inducing Rabi oscillations between the states.

\newcommand{\tabsz}{\small}
\begin{table}[b]
  \begin{tabular}{|m{1.2cm}|m{2.35cm} |m{2.35cm} |m{2cm}|}
\hline
{\bf Class }&  {\bf Dissipative}  & {\bf Unitary}  & {\bf Dispersion}\\
\hline \large
\begin{minipage}{1.2cm} \tabsz $I_s$ ($k_0\neq0$, $\omega_0=0$)\end{minipage} &  
\tabsz Eckhaus instability in convection; Turing instability in morphogenetics &  
\tabsz Mode softening in 2D dipolar condensates and spinor condensates &
 \show{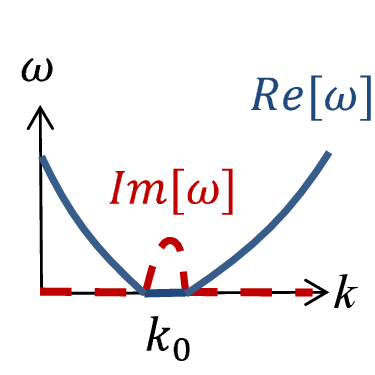} \\
\hline
\begin{minipage}{1.2cm}\tabsz $III_o$ ($k_0=0$, $\omega_0\neq0$)\end{minipage} &  
\tabsz Belousov and Zhabotinsky (BZ) chemical reactions & 
\tabsz Pairing instability near Feshbach resonances & \show{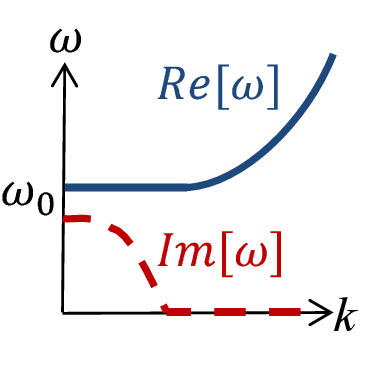} \\
\hline
\begin{minipage}{1.2cm}\tabsz $I_o$ ($k_0\neq0$, $\omega_0\neq0$)\end{minipage} & 
\tabsz Convection in binary mixtures  & 
{\it \tabsz Unstable avoided crossing in spinor condensates }& \show{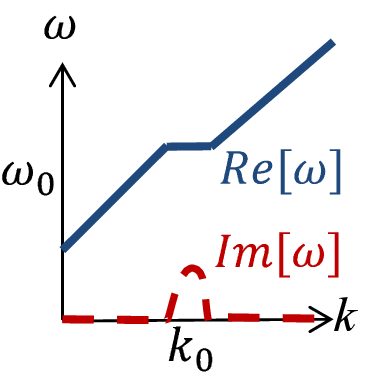} \\
\hline
\end{tabular}
\caption{Classification of dynamical instabilities as static (s) and oscillatory (o), with examples of dissipative (classical) and unitary (quantum) systems. The focus of the present work is in the quantum domain of class $I_o$. } 
\label{tab:classification}
\end{table}

A system of this type was recently realized by Nicklas {\it et al.} \cite{Oberthaler}, with two hyperfine states of $^{87}\mathrm{Rb}$ coupled by a field close to atomic resonance. 
  In this experiment, the initial state corresponded to a fully polarized condensate and the dynamics was induced by suddenly switching on the driving field.
   In contrast, we consider here the effects of a quench in the phase of the field. 
   At a mean-field level, the system jumps from a minimum to a maximum of the energy and can remain dynamically stable. By considering the Bogoliubov excitations around this state \cite{Tommasini}, we find that instabilities necessarily arise around specific wave vectors, determined by an interplay between kinetic energy, coherent pump, and interactions.

\paragraph*{} 
To develop insight into the different types of pattern formation, we briefly review the stability conditions of quadratic modes in unitary quantum systems. Small fluctuations around a given initial state can be energetically stable, dynamically stable, or unstable. 
In the case of a single degree of freedom, the energetically stable mode
can be illustrated by a harmonic oscillator with Hamiltonian
$H=\frac{1}{2}\omega ( p^2 + x^2 )$.
The mode is characterized by a ground state of minimal energy and a real frequency $\omega$, the energy quantum of excitation.
In contrast, the dynamically stable mode is exemplified by the Hamiltonian
$H=-\frac{1}{2} \omega ( p^2+x^2 )$.
The frequency is still real, but the vacuum state has now maximal energy.
The dynamics of energetically and dynamically stable modes are identical, but the latter cannot represent thermal equilibrium.
Finally, a mode can be dynamically unstable.
The simplest case is an inverted harmonic potential, with
$H=\frac{1}{2} \omega ( p^2-x^2 )$. 
The dynamics is described by exponentially growing fluctuations, rather than by periodic oscillations, and can be associated with an imaginary frequency $i\omega$
\footnote{The analogy between dynamically unstable modes and imaginary frequencies can be made explicit by the transformation $p'=\sqrt{i}p,~x'=\sqrt{i}x$, satisfying $\omega(p^2-x^2)=i\omega (x'^2+p'^2)$}.
As will be illustrated below, this notion can be generalized to complex frequencies with both real and imaginary parts: the states will have both oscillatory and exponentially growing dynamics.

We now consider the fluctuations of a many-body translation-invariant system. 
In this case, a collective mode is represented by a sum of harmonic oscillators
$H = \sum_k \pm \frac{1}{2}  \omega (k) (p_{-k} p_k \pm x_{-k} x_k)$,
where the three stability scenarios are possible for each value of the wave vector $k$.
In the case of interest, we consider two coupled modes described by the Hamiltonian
\begin{equation}
  H= \sum_k \frac{1}{2} 
  \ve p^\dagger
  M_p
  \ve  p
  + \frac{1}{2}
  \ve x^\dagger
  M_x
  \ve x
  \label{eq:Hquadmatrices}
\end{equation}
with $M_p=M_p(k)$, $M_x=M_x(k)$, 
${\bf p}=\big(p_{1,k},~p_{2,k}\big)^\top$, and  
${\bf x}=\big(x_{1,k},~x_{2,k}\big)^\top$.

To diagonalize the system and find the Bogoliubov modes, we use a dynamical method similar to the formalism of Rossignoli and Kowalski \cite{Rossignoli}.
We first compute the  equations of motion of the canonical operators $x_{i,k}$ and $p_{i,k}$ in the Heisenberg picture
\begin{equation}
  \frac{d}{dt} \MAT{\ve x \\ \ve p  } = 
  \MAT{\mathbb 0 & M_p \\  -M_x & \mathbb 0 } \MAT{\ve x \\ \ve p  }.
  \label{eq:HeisenbergEOM}
\end{equation}
We then define the Bogoliubov operators as the linear combination 
$b_{i,k}= a_1 x_{1,k} + a_2 x_{2,k} + a_3 p_{1,k} + a_4 p_{2,k}$ and require them to be  eigenoperators of the Heisenberg equation of motion 
$i \frac{d}{dt} b_{i,k} = \omega_{i,k} b_{i,k}$.
Going to the second time derivative, we finally obtain
\begin{equation}
   \ve a^\top \omega_{i,k}^2= 
  \ve a^\top
   \MAT{M_p M_x & \mathbb 0 \\   \mathbb 0 & M_x M_p }
   \label{eq:frequencies}
\end{equation}
with $\ve a^\top = (a_1, a_2, a_3, a_4)$.
Finding the Bogoliubov operators and their frequencies has reduced to solving a block diagonal $4\times4$ linear system. The frequencies $\omega_{i,k}$ are given by the square root of the eigenvalues of 
$M_xM_p$ or $M_pM_x$.

Knowing how to diagonalize systems of the form (\ref{eq:Hquadmatrices}), we introduce a generic model of level crossing with
\begin{equation}
  M_p =
  \MAT{\omega_1(k) &0 \\ 0 & \omega_2(k)},
  \quad
  M_x =
  \MAT{\omega_1(k) & \varepsilon \\ \varepsilon & \omega_2(k)}\;,
\end{equation}
where $\varepsilon$ is some coupling introduced between the two modes. For $\varepsilon=0$, the two modes are decoupled and the eigen-frequencies of the system are simply $\omega_1(k)$ and $\omega_2(k)$. If $\omega_i(k)>0$, the mode is energetically stable, while if $\omega_i(k)<0$ it is only dynamically stable.
For $\varepsilon\neq0$, the original modes are mixed and 
we find the Bogoliubov frequencies by diagonalizing $M_x M_p$ for each wave vector $k$: 
\begin{equation}
  \omega^2(k) = \frac{\omega_1^2 + \omega_2^2}{2}
  \pm \sqrt{
  \left(\frac{\omega_1^2 - \omega_2^2}{2}\right)^2
  + \varepsilon^2 \omega_1 \omega_2
  }.
  \label{eq:toyfreqs}
\end{equation}
At a level crossing, defined by $|\omega_1|=|\omega_2|$, the squared frequency difference is given by 
$\sqrt{\varepsilon^2 \omega_1 \omega_2}$.
If both modes are energetically stable ($\omega_1,\omega_2>0$), the two eigen-frequencies never coincide, and one obtains the usual avoided level crossing (Fig.\ \ref{fig:levelavoidance}(a)). In contrast, if one mode is energetically stable ($\omega_i>0$) and the other only dynamically stable ($\omega_j<0$),
the frequency difference becomes complex and the system is unstable with a finite real component to the frequency (Fig.\ \ref{fig:levelavoidance}(b)). 

\paragraph*{}  
An experimental system in which the above-mentioned effect can be observed is a condensate with two internal degrees of freedom \cite{Ho96,Stenger98,Burke98,Pethick&Smith,Hall08,Molamed08} (states $\Ket 1$ and $\Ket 2$, separated by the energy $\hbar \omega$ )
and coherently coupled by a radio-frequency (RF) field
\footnote{
We note that in the experiment, two-photon processes can be used, with the RF field combined with microwave radiation for instance \cite{Oberthaler}, with the same result.
}.
The RF coupling contribution to the Hamiltonian is
$H_{\mathrm{RF}} = -\alpha ( e^{i\nu t} \OPd 1 \OP 2 + e^{-i\nu t} \OPd 2 \OP 1)$,
with $\alpha>0$ the Rabi frequency, $\nu$ the frequency of the field and
$\OPd 1$ and $\OPd 2$ bosonic creation operators for states $\Ket1$ and $\Ket 2$.
In the interaction frame with respect to 
$H_0 = \hbar \nu (\OPd 2 \OP 2 - \OPd 1 \OP 1)$,
time-dependence is cancelled and the Hamiltonian becomes
\begin{align}
    \hat H =& \int dx \; 
     \sum_i  -\OPd i \frac{ \derx^2}{2m} \OP i + \frac{\lambda_i}{2} \OPd i \OPd i \OP i \OP i + \lambda_{12} \OPd1 \OPd2 \OP2 \OP1\nonumber \\
     &+\frac{\delta}{2} \left( \OPd 2 \OP 2 - \OPd 1 \OP 1 \right) 
      - \alpha \left( \OPd1 \OP2 + \OPd2 \OP1 \right),
  \label{eq:Hamiltonianprincipale}
\end{align}
where we set $\hbar=1$,
$m$ is the mass of the atoms and
$\delta=\omega-\nu$ the detuning.
The parameters $\lambda_1$, $\lambda_2$, and $\lambda_{12}$ describe state-dependent local interactions, which can be tuned with the aid of a Feshbach resonance (see Ref.~\cite{Cheng10} for a review).

The equilibrium properties of the Hamiltonian (\ref{eq:Hamiltonianprincipale}) are well understood \cite{Tommasini}. In the absence of the coherent drive ($\alpha=0$), the population of each state is fixed and the system undergoes a miscible-immiscible phase transition, depending on the relative strength of the interactions\cite{Papp08}. If $|\lambda_{12}| < \sqrt{\lambda_1 \lambda_2}$, the gas is miscible, i.e. it is energetically favorable to have both species share space rather than creating a phase separation. 
Alternatively, if $\lambda_{12}>\sqrt{\lambda_1 \lambda_2}$, the gas is immiscible: the phase-separate state has the lowest energy. Finally, if $\lambda_{12}<-\sqrt{\lambda_1 \lambda_2}$, the system collapses. We note that the phase transition corresponds to the appearance of a dynamical instability. For example, when a miscible gas becomes immiscible, waves of population imbalance (``spin mode'') become unstable at long wavelength and the fluctuations tend to move the  system towards the new phase-separated equilibrium.

In the presence of a coherent drive, the mean-field phase diagram of the Hamiltonian (\ref{eq:Hamiltonianprincipale}) can be found by
replacing the field operator by a classical field
$\OP i \approx \OPz i = \sqrt{n_i}e^{-i\theta_i}$ and looking for stationary solutions 
$\OPz i (t) = \OPz i (0) e^{-i\mu t}$
to the Gross-Pitaevski equations \cite{Pethick&Smith}. This approach leads to two main families of solutions, characterized by the phase difference $\theta=\theta_1 - \theta_2$: when $\theta=0$ the coupling energy is minimal and when $\theta=\pi$ it is maximal. For each value of $\theta$, there can be up to three stationary solutions corresponding to different values of the population imbalance $n_1-n_2$ 
\footnote{ The population imbalance $f = (n_1-n_2)/(n_1+n_2)$ is given \cite{Tommasini} by the solution of  
\begin{equation*}
  \delta + (\lambda_1-\lambda_2) n
  + (\frac{\lambda_1 + \lambda_2}{2} - \lambda_{12})n f 
  + \frac{2 \alpha \cos \theta f}{\sqrt{1-f^2}} 
  =0
  \label{eq:fequation}
\end{equation*}
with $n=n_1 + n_2$ and can have up to three possible values. In the symmetric case $\delta=0$ and $\lambda_1=\lambda_2$ one solution always corresponds to $f=0$. See also Tommasini {\it et al.} \cite{Tommasini} for a discussion of the solutions around $\theta=0$ and the nature of the Bogoliubov expansion around it.}. 
Here we consider only the symmetric solution $n_1=n_2$, which leads to simpler analytical results and can be easily prepared in experiments, as explained below.

To study the stability of the system, we now consider small quantum fluctuations by approximating
$\OP i \approx \OPz i + e^{i \theta_i} \sum_{k\neq0} \frac{e^{i k \cdot x}}{\sqrt{V}}a_{i,k}$,
 in (\ref{eq:Hamiltonianprincipale}) and expanding up to second order.
Going to a frame that rotates at the same frequency as the condensate,
we find that the linear term cancels and we are left with a quadratic bosonic system.
For convenience, we introduce the canonically conjugated variables 
$x_{i,k} = (a_{i,-k}^\dagger - a_{i,k})/i \sqrt{2}$ and 
$ p_{i,k} = (a_{i,-k}^\dagger + a_{i,k})/{\sqrt{2}}$
which are respectively linked to phase and density fluctuations \footnote{
As can be seen by
$x_i = \sqrt{2n_i} \dPO i$ and
$p_i =\dDO i / \sqrt{2n_i}$
that we obtain if we write
$\OP i = e^{-i(\theta_1 + \dPO i)} (n_i + \dDO i)^{1/2}$
and expand it to first order.
}.
In this basis, the system is described by a Hamiltonian of the form (\ref{eq:Hquadmatrices})
with the matrices $M_p$, $M_x$ now given by
\begin{align}
  M_p &=
 2
 \begin{pmatrix}
   \lambda_1 n_1 & \lambda_{12} \sqrt{n_1 n_2} \\
   \lambda_{12} \sqrt{n_1 n_2} & \lambda_2 n_2
 \end{pmatrix}
 + \alpha \cos \theta
 \begin{pmatrix}
    \sqrt{\frac{n_2}{n_1}} & -1 \\
    -1 & \sqrt{\frac{n_1}{n_2}}
 \end{pmatrix}\nonumber\\
  &+ \frac{k^2}{2m} 
  \begin{pmatrix}
    1 & 0 \\ 0 & 1
  \end{pmatrix}
  + \frac{\delta}{2}
  \MAT{-1&0\\0&1},\\
  M_x &=
  \alpha \cos \theta
  \begin{pmatrix}
    \sqrt{\frac{n_2}{n_1}} & -1 \\
    -1 & \sqrt{\frac{n_1}{n_2}}
  \end{pmatrix}
  + \frac{k^2}{2m} 
  \begin{pmatrix}
    1 & 0 \\ 0 & 1
  \end{pmatrix}
  + \frac{\delta}{2}
  \MAT{-1&0\\0&1}.
  \label{eq:matrixMx}\nonumber
\end{align}
As explained above, the Bogoliubov frequencies are  the square roots of the eigenvalues of $M_x M_p$.

\begin{figure}[t]
  \begin{center}
   \centering
  \includegraphics[width=\columnwidth]{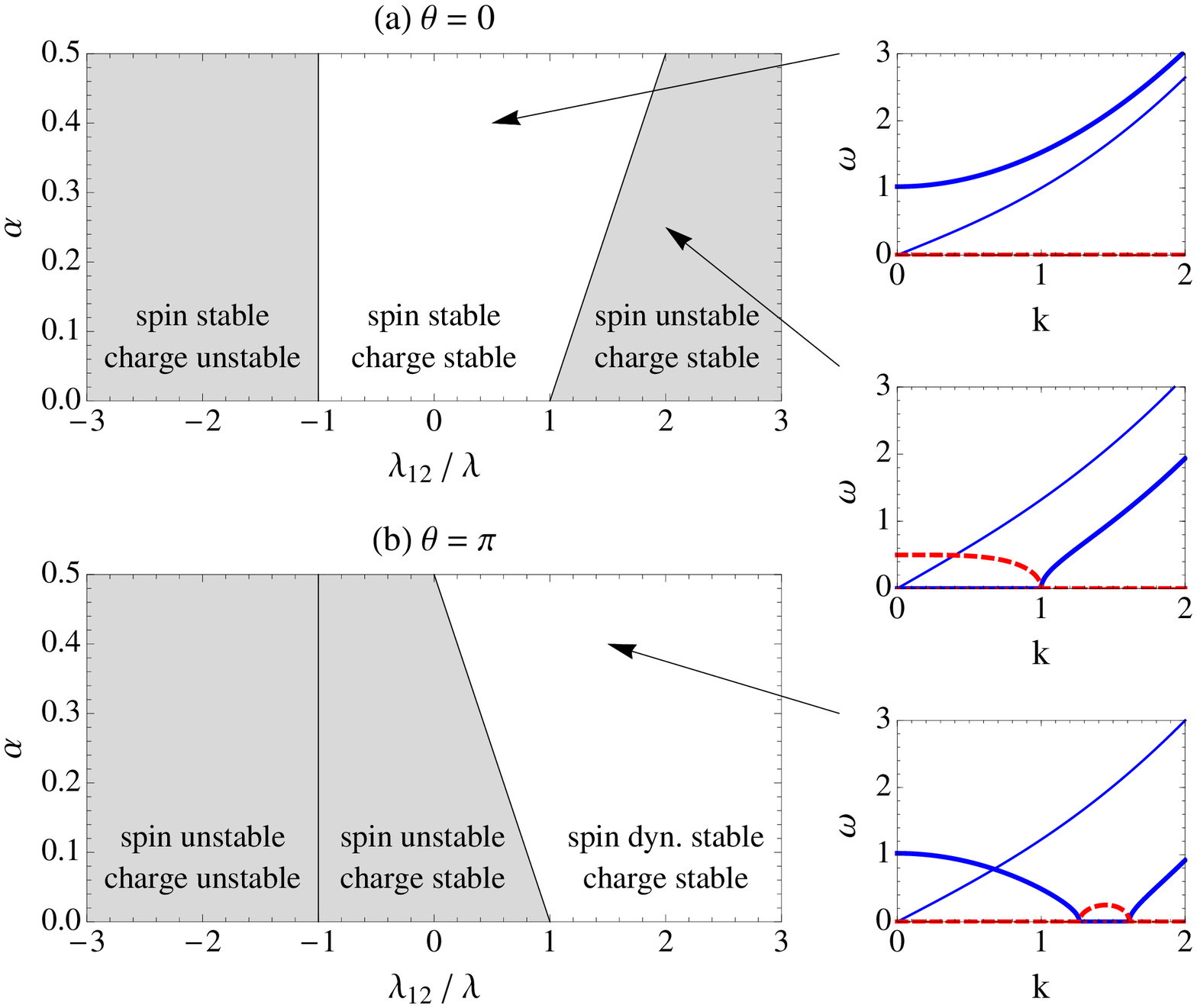}

  \end{center}\vspace{-0.5cm}
  \caption{(color online).
  Stability diagrams of the system for the symmetric case  
  ($\lambda=\lambda_1=\lambda_2$, $n_1=n_2$, $\delta=0$), 
  with solutions (a) $\theta=0$ (equilibrium) and (b) $\theta=\pi$ (non-equilibrium). 
  Grey background indicates dynamical instability at small wave vectors.
  The insets illustrate the real (blue solid line) and imaginary (red dashed line) parts of the dispersions of the two modes (thinner lines for the charge mode and thicker for the spin mode). The variables are in dimensionless units with $m=1$, $\hbar=1$ and the energy measured in $\lambda n$.
  }
  \label{fig:Diagtheta}
\end{figure}

We first consider the symmetric case $\lambda=\lambda_1 = \lambda_2$ and $\delta=0$, with the mean-field solutions $n_1=n_2$ and $\theta=0,\pi$. In this case, $M_x$ and $M_p$ have the same eigenvectors for all $k$ and the system decouples into
a charge mode
($\sqrt{2} x_c = x_1 + x_2 $, $\sqrt{2} p_c = p_1 + p_2$)
and a spin mode
($\sqrt{2} x_s = x_1 - x_2$, $\sqrt{2} p_s = p_1 - p_2$).
For the charge mode, the Hamiltonian is
\newcommand{\kinE}{\frac{k^2}{2 m}}
\begin{equation}
  H_c = \frac{1}{2} \left(
  \Big(n (\lambda + \lambda_{12}) + \kinE \Big) p_c^2
  +
  \Big( \kinE \Big) x_c^2
  \right)
  \label{eq:HamiltCharge}
\end{equation}
which does not depend on the coupling $\alpha$ nor the phase $\theta$.
The condition for stability is that both terms are positive, or equivalently $\lambda_{12} > - \lambda$.
For the spin mode, we obtain
\begin{equation} 
  \begin{split}
  H_s = \frac{1}{2} \bigg(&
  \Big(n (\lambda - \lambda_{12}) + 2 \alpha \cos \theta + \kinE \Big) p_s^2\\
  &+
  \Big( 2 \alpha \cos \theta + \kinE \Big) x_s^2 
  \bigg)
  \end{split}
  \label{eq:HamiltSpin}
\end{equation}
and both the coupling strength $\alpha$ and the phase $\theta$ will affect the stability of the system.
For $\alpha=0$, the stability condition reduces to the miscibility condition 
$\lambda_{12} < \lambda$.
For $\alpha\neq0$ and the equilibrium solution $\theta=0$, this condition is simply extended to 
$\lambda_{12} < \lambda+ 2 \alpha/n$ (see Fig.\ \ref{fig:Diagtheta}(a)).

Things get more interesting for the non-equilibrium solution $\theta=\pi$. In this case, the coefficient in front of $x_s^2$ in Eq.~(\ref{eq:HamiltSpin}) is always negative for $k$ small enough.
The system can nevertheless become {\em dynamically} stable when the coefficient in front of $p_s^2$ is negative as well,
translating to the condition $\lambda_{12} > \lambda - 2 \alpha/n$ (see Fig.\ \ref{fig:Diagtheta}(b)).
We note that the system is stable for large $\lambda_{12}$ in contrast to the uncoupled case, where the system was stable for small $\lambda_{12}$. This is the result of a competition between the interactions and the coherent coupling, compensating each other to avoid the instability.
For larger wave vectors, the kinetic energy
$k^2/(2m)$ 
will make both coefficients less and less negative until they switch sign.
Except for the special case $\lambda_{12}=\lambda$ (where both coefficients are equal) there will be an unstable interval, 
when one coefficient is positive and the other negative. This gives rise to a dynamical instability of type $I_s$, not directly related to mode softening, shown in Fig.\ \ref{fig:Diagtheta} (third inset). 

\begin{figure}[t]
  \begin{center}
  \newlength{\halffigurewidth}
  \setlength{\halffigurewidth}{0.23\textwidth}
    \SUBFIG{
    \includegraphics[width=\halffigurewidth]{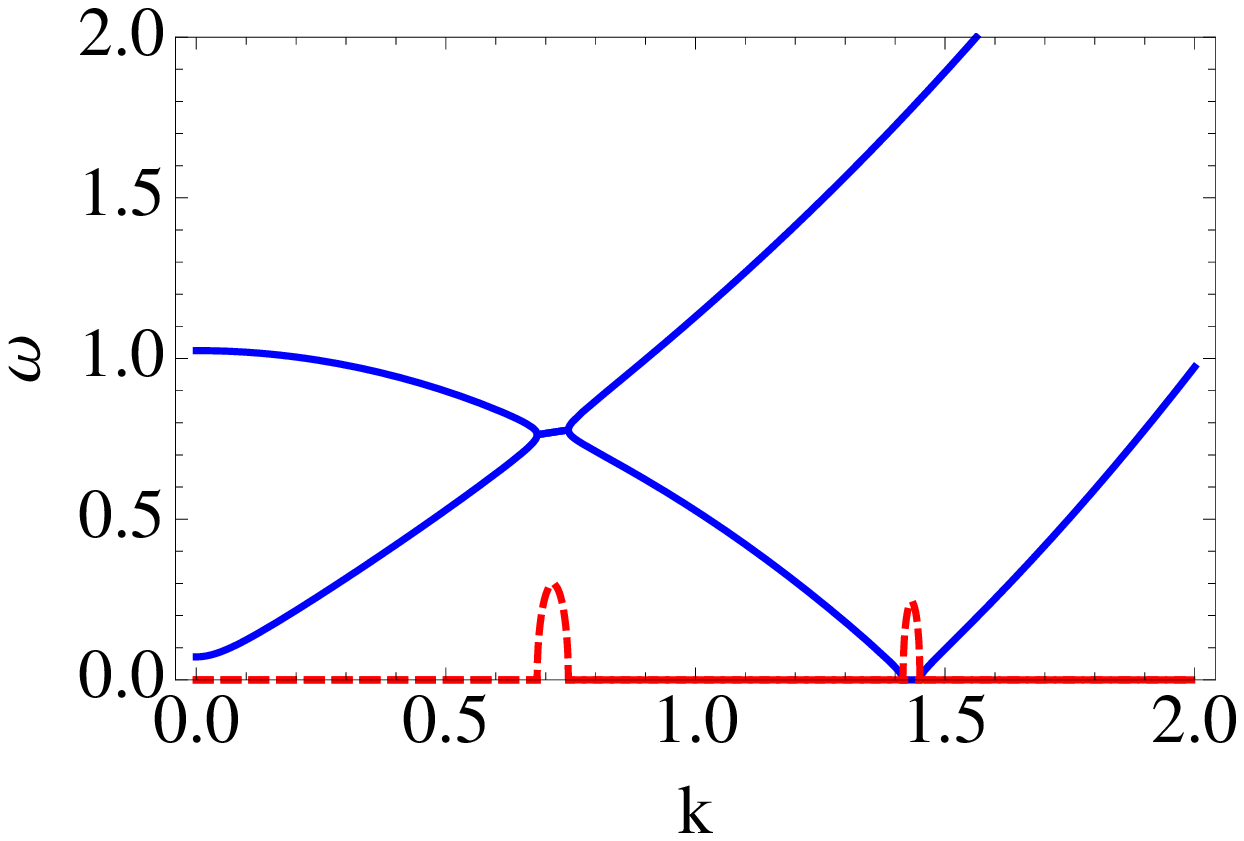}
\label{subfig:newinstab}}
    \SUBFIG{
    \includegraphics[width=\halffigurewidth]{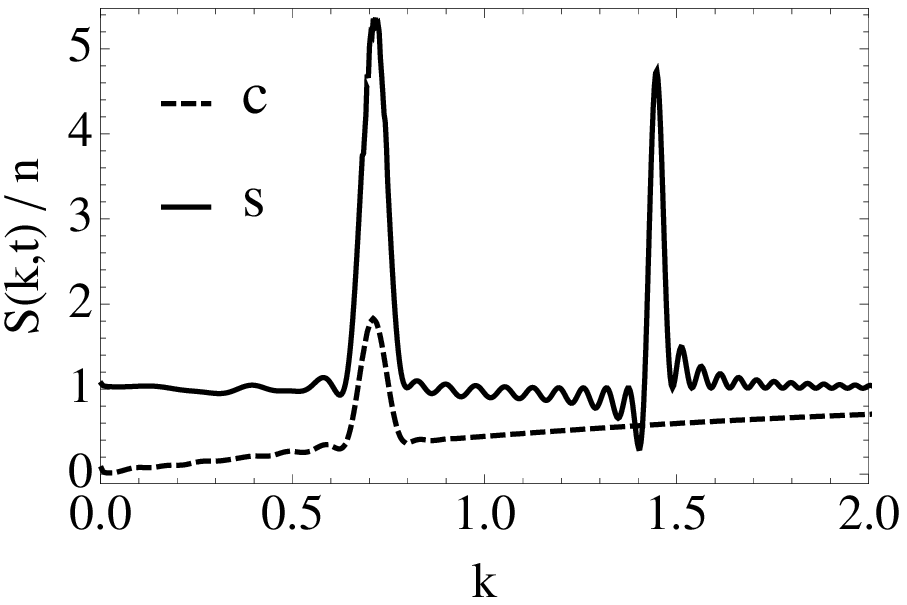} 
\label{subfig:newinstab_power}}
  \end{center}
\vspace{-0.5cm}
  \caption{(color online).
 \protect\subref{subfig:newinstab} Dispersion relation in the asymmetric case $\delta\neq0$, showing two dynamical instabilities (see text). 
  The real part of the frequency is the solid blue line and the imaginary part (multiplied by 10 for better legibility) is the dashed red line.  
  \protect\subref{subfig:newinstab_power}
Power-spectrum of the charge (dashed line) and spin (solid line) sectors, showing sharp peaks in Fourier space where the dispersion relation displays instabilities. 
  Numerical parameters: $\lambda_{12} = 1.05 \lambda$, $\alpha = 0.5 \lambda n$, and $\delta = 0.1 \lambda n$ and  $t=40.0 \hbar/\lambda n$.  
  The variables are in dimensionless units with $m=1$, $\hbar=1$ and the energy measured in $\lambda n$.
  }

  \label{fig:Predictions}
\end{figure}

If we depart from the symmetric case, by considering $\lambda_1\neq\lambda_2$ or $\delta\neq0$,
the matrices $M_x$ and $M_p$ will in general no longer have the same eigenvectors and the perturbation will mix the charge and spin modes.
If the deviation from the symmetric case is not too large, the effect on the dispersion will be small, except at energy crossings.
In our system we observe an energy crossing only for the non-equilibrium solution $\theta=\pi$, where the spin mode, first decreasing in frequency, crosses the charge mode, as shown in the third inset of Fig.\ \ref{fig:Diagtheta}.
As explained above, a novel instability will be introduced as the crossing between the energetically stable charge mode and the dynamically stable spin mode is avoided (see Fig.~\ref{fig:Predictions}\subref{subfig:newinstab}).
Up to the first order in 
perturbation theory,
we find the instability to be be centered around
\begin{align}
  k_0 &=
  \sqrt{
  2 m \alpha
  \frac{2 \alpha + (\lambda_{12} - \lambda)n}{2 \alpha +  \lambda_{12} n}
  },
  \label{eq:konset}\\
  \omega_0 &=
  \sqrt{\frac{
  \alpha (\alpha + \lambda_{12} n) 
  \left( 4 \alpha ( \alpha + \lambda_{12}n ) + (\lambda^2 - \lambda_{12}^2)n^2 \right)}
  {(2 \alpha + \lambda_{12}n)^2}}.
  \label{eq:omonset}\nonumber
\end{align}
These quantities depend in a non-trivial way on the coherent coupling and on the interactions, signaling that it results from interplay and competition between the two.

\paragraph*{} 
We now describe an experimental protocol designed to probe the interesting properties of the $\theta=\pi$ solution and its instabilities.
In order to achieve the relevant state, we propose to apply a finite coherent drive $\alpha$ and a large detuning $\delta$,
and start with the atoms in the ground state $\Ket 1$.
If the detuning $\delta$ is adiabatically reduced to a small finite value, the system will end up in the minimal-energy solution $\theta=0$. We then propose to quench the system by changing $\alpha\to-\alpha$
(by introducing a phase shift of $\pi$ in the coupling-inducing field),
equivalent to  changing $\theta=0\to\pi$.
The system is now in the appropriate mean-field state 
($\theta=\pi$, $n_1 \approx n_2$) that we let evolve for a time $t$ before proceeding to a measurement.

Due to quantum fluctuations, the unstable modes of the quenched Hamiltonian will develop and grow exponentially.
Charge and spin density waves will form at the unstable wave vectors, creating regular patterns at the associated wavelengths.
Because the position of these patterns varies from one realization to another, the average density will not be affected by the dynamical instabilities.
Nevertheless, the existence of unstable modes can be probed by measuring the power-spectrum of the fluctuations in the local density of atoms\cite{Cherng08,Vengalattore08}, defined by 
\begin{equation} S_{s/c}(k,t)=\langle \big|\delta\rho_{s/c}(k,t)\big|^2\rangle 
\;, \end{equation}
where 
$\delta\rho_c(k)=\int dx ~\frac{e^{-i k x}}{\sqrt{V}} \big(\OPd 1(x)\OP 1(x)+\OPd 2(x)\OP 2(x)-n\big)$ and 
$\delta\rho_s(k)=\int dx ~\frac{e^{-i k x}}{\sqrt{V}} \big(\OPd 1(x)\OP 1(x)-\OPd 2(x)\OP 2(x)\big)$ are respectively the momentum-resolved charge (c) and spin (s) fluctuations. 
Within the approximations of the present analysis (i.e. small fluctuations), the power spectrum is simply given be 
$S_{s/c}(k,t) = 2n\langle p_{s/c}(k,t)p_{s/c}(-k,t)\rangle$ and plotted in Fig.~\ref{fig:Predictions}\subref{subfig:newinstab_power}.
As expected, both $S_c(k,t)$ and $S_s(k,t)$ show sharp features at the wave vectors corresponding to the dynamical instabilities predicted by the dispersion relation shown in Fig.~\ref{fig:Predictions}\subref{subfig:newinstab}. 
In particular, the charge mode displays only one dynamical instability at $k\approx0.7$, induced by the unstable avoided crossing with the (otherwise dynamically stable) spin mode 
\footnote{
See the Supplemental Material for an alternative experimental method to probe the predicted instability. 
}.

\paragraph*{} 
In summary, we have found that the dynamics of coherently-driven
spinor condensates reveals interesting properties in the intermediary regime where the coherent coupling is comparable to the interaction energy. 
By quenching the phase of the coherent drive it is possible to induce a dispersion relation featuring an energy crossing between an energetically-stable  charge mode and a dynamically-stable spin mode.
Quite generically, such a crossing becomes unstable when a perturbation is introduced to mix the modes.
For our system, this can be easily obtained by introducing a finite detuning of the coupling field frequency.
We propose an experimental protocol designed to test our result and predict the consequences of the instability.

This mechanism, which we term unstable avoided crossing, is a generic feature of systems with two degrees of freedom in a  non-equilibrium state.
We note that in traditional condensed-matter systems, negative frequencies are often associated with dynamical instabilities 
(for instance in the case of the Landau instability in superfluids).
Using our results, we can understand why this is the case.
In the presence of a dissipative bath, any negative-frequency mode can couple to a stable mode of the bath with the same frequency (in absolute value), leading to  dynamical instability.
In contrast, for ultracold neutral atoms, the gas is extremely well insulated from the environment and negative-frequency modes are dynamically stable.

There are multiple pathways to extend this work. In particular, we described the system assuming Bose-Einstein condensation. This approach does not apply to one-dimensional systems, which should rather be modeled as quasi-condensed Luttinger liquids. 
One might study whether an avoided level crossing still arises within this treatment.
A second possible research direction regards the effect of optical lattices, which effectively introduce a momentum cutoff with the Brillouin zone. 
By changing the density of the atoms, it should be possible to limit the maximal value of the allowed momentum and, in this way, control the emergence of finite-wave vector dynamical instabilities.

We would like to thank F.\ Mila, I.\ Carac, M.\ Oberthaler and R.\ Barnett for useful discussion. The authors acknowledge support from Harvard-MIT CUA, DARPA OLE program, ARO-MURI on Atomtronics, ARO-MURI Quism.

\bibliography{bibli}

\end{document}